\documentclass[10pt,twocolumn,letterpaper]{article}

\usepackage{cvpr}      

\usepackage{graphicx}
\usepackage{amsmath}
\usepackage{amssymb}
\usepackage{booktabs}

\DeclareMathOperator*{\argmin}{arg\,min}
\usepackage{multirow,bigstrut}
\usepackage{overpic}
\usepackage{subcaption}
\usepackage{color}
\newcommand{\wdz}{\textcolor{black}}
\newcommand{\hyy}{\textcolor{black}}
\newcommand{\cvpr}{\textcolor{black}}
\newcommand{\HYY}{\textcolor{black}}
\usepackage[pagebackref,breaklinks,colorlinks]{hyperref}

\usepackage[capitalize]{cleveref}
\crefname{section}{Sec.}{Secs.}
\Crefname{section}{Section}{Sections}
\Crefname{table}{Table}{Tables}
\crefname{table}{Tab.}{Tabs.}

\def\cvprPaperID{4527} 
\def\confName{CVPR}
\def\confYear{2022}

\makeatletter

\newcommand{\Rmnum}[1]{\expandafter\@slowromancap\romannumeral #1@}
\makeatother

\begin{document}

\title{Neural Data-Dependent Transform for Learned Image Compression}

\author{Dezhao Wang \qquad Wenhan Yang \qquad Yueyu Hu \qquad Jiaying Liu \\
Wangxuan Institute of Computer Technology, Peking University \\
\small{\texttt{\{wangdz, yangwenhan, huyy, liujiaying\}@pku.edu.cn}} \\
}
\maketitle

\begin{abstract}

Learned image compression has achieved great success due to its excellent modeling capacity, but seldom further considers the Rate-Distortion Optimization (RDO) 
\wdz{of each input image.} 
To explore this potential in the learned codec, we make the first attempt to build a neural data-dependent transform and introduce a continuous online mode decision mechanism to jointly optimize the coding efficiency for each 
\HYY{individual image}. Specifically, apart from the image content stream, we employ an additional model stream to generate the transform parameters at the decoder side. 
The presence of a model stream enables our model to learn more abstract neural-syntax, which helps cluster the latent representations of images more compactly. 
\cvpr{Beyond the transform stage, we also adopt neural-syntax based post-processing for the scenarios that require higher quality reconstructions regardless of extra decoding overhead.} Moreover, the involvement of the model stream further makes it possible to optimize both the representation and \HYY{the} decoder in an online way, \textit{i.e.} RDO at the testing time. It is equivalent to a continuous online mode decision, like coding modes in the traditional codecs, to improve the coding efficiency based on the individual input image. The experimental results show the effectiveness of the proposed neural-syntax design and the continuous online mode decision mechanism, demonstrating the superiority of our method in coding efficiency \cvpr{compared to the latest conventional standard Versatile Video Coding (VVC) and other state-of-the-art learning-based methods}. Our project is available at: \url{https://dezhao-wang.github.io/Neural-Syntax-Website/}.

\end{abstract}

\section{Introduction}

{

Image compression is one of the most fundamental technologies since human society has entered the digital information age.
It becomes more and more important when currently the big data applications meet the constantly increased visual experience \hyy{demand}, \textit{e.g.} high-resolution visual applications like 8K streaming and Virtual Reality~(VR).
Continuous endeavors are made to obtain highly efficient compressed and high-quality images/videos that can be stored, displayed, and analyzed with limited hardware resources.

In recent decades, a series of codecs have been developed to optimize the reconstruction quality (\textit{i.e.} distortion) with bit-rate constraints (rate) jointly, which forms the core problem of lossy compression: \textit{rate-distortion optimization} (\textit{RDO}).
It perfectly describes the key aspects of human needs for a large amount of images/videos: maximizing reconstruction quality leads to preserving the critical visual information of the image signal;
whereas minimizing the bit-rate benefits the efficient transmission and storage.

\begin{figure}[t]
    \centering
    \includegraphics[width=\linewidth]{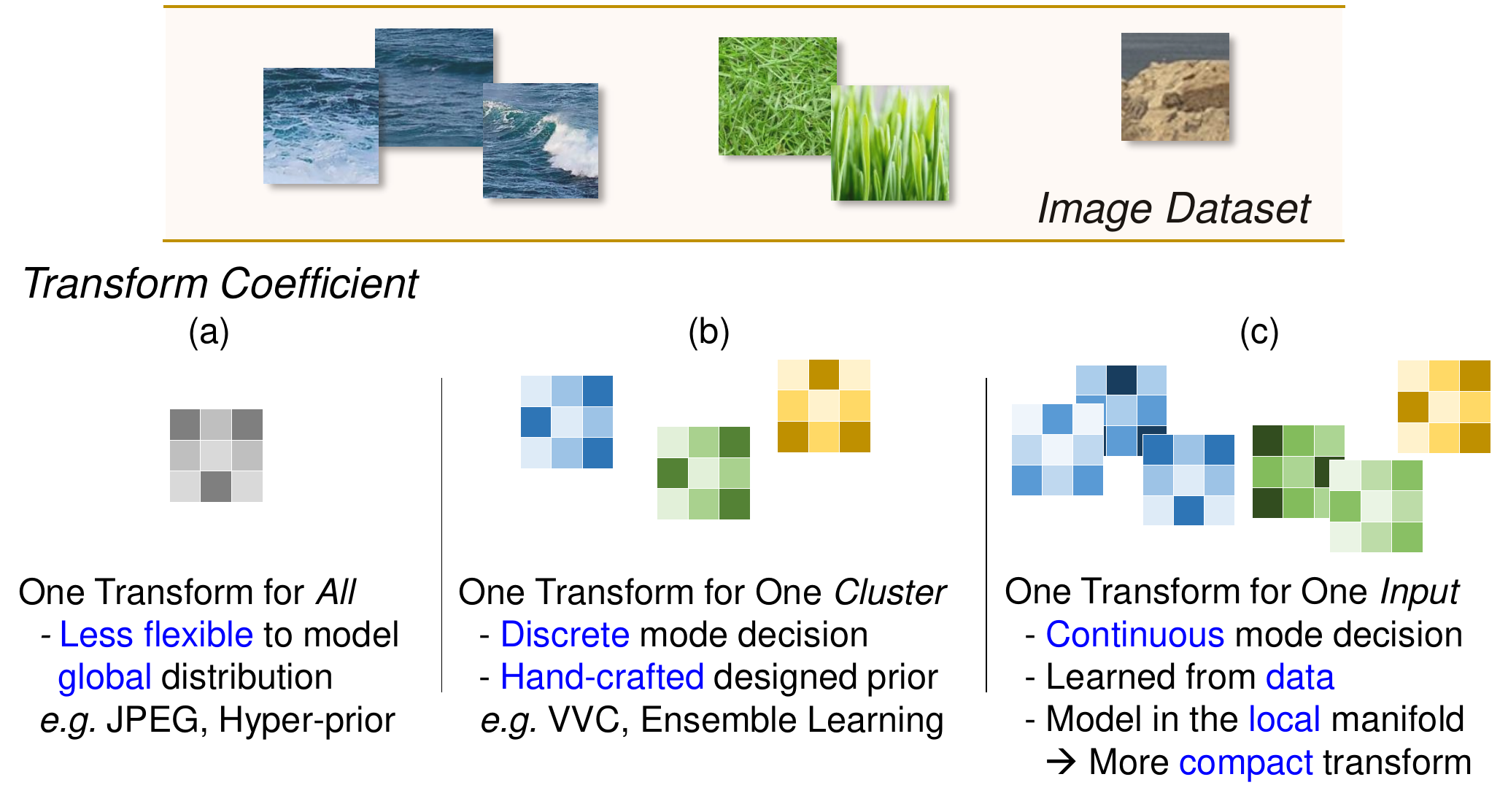}
    \caption{
    Three transform paradigms: 
    (a) Single transform adopted in previous end-to-end trained image compression methods;
    (b) Cluster-specified transform utilized in conventional image codecs;
    (c) \wdz{Proposed data-dependent transform}.
    In our method, the transform is generated based on the input sample.
    By modeling the distribution in a local manifold, we derive a more compact transform and enable the decoder to be more powerful and flexible.
    }
    \vspace{-4mm}
    \label{fig:teaser}
\end{figure}

The mainstream image compression standards and systems, \textit{e.g.} JPEG~\cite{wallace1992jpeg} and BPG~\cite{bpg} based on HEVC~\cite{sullivan2012overview}, adopt the route of transform/hybrid coding framework for RDO.
The framework consists of cascaded transform, quantization and entropy coding.
The elaborate designs of these components lead to higher coding efficiency.
Among these efforts, there are two important directions. 
One branch of researches focuses on designing more powerful transforms,
\hyy{\textit{e.g.} the improved variants~\cite{RABBANI20023,DST_intra_pred,mode_dependent} of discrete cosine transform (DCT)~\cite{bpg,vvc}, and the theoretically optimal linear Karhunen-Lo{\`e}ve transform (KLT)~\cite{zhang2020image}.}
Although more decorrelated and \wdz{energy-compact} coefficients are obtained to improve the coding performance,
\hyy{these methods heavily rely on the distribution, and therefore are not general and flexible enough}.
The other branch of works pay attention to fully capturing the properties of the input samples. These methods introduce syntax elements to project the image signals into a specific subspace, \textit{e.g.} \wdz{intra}-prediction based on different directions,
where more compact representations can be naturally obtained and related processing can be more data-dependent. However, the syntax and corresponding subspace are \wdz{manually} predefined, \textit{e.g.} the directional modes of intra-prediction, which lead to limited performance gains and leave less space for future improvement.

With the rapid development of deep learning, the prosperity of end-to-end optimized image compression methods is witnessed.
In these methods, the whole neural codecs~\cite{balle2015density,balle2016end} (including encoders and decoders) are totally learned from a large collection of high-quality images.
Via optimizing the rate-distortion (R-D) cost over the large-scale training set, the encoders provide flexible and powerful nonlinear neural transforms.

However, existing end-to-end optimized image compression methods 
\wdz{seldom pay attention}
to the model's adaptivity to handle \hyy{images with diverse} contexts or distributions. 
The training only leads to an average low R-D cost on the training set. 
For a given input sample, the codec might not be good at capturing the input's 
\HYY{probability properties} 
and fail to provide an optimal transform during the inference stage.
Some insights in traditional codecs bring in new inspirations.
First, transforms can be totally data-dependent~\cite{Variable_block_size,Annealed,SVD} instead of using fixed weights.
Second, \textit{syntax} might be very useful to simplify the distribution of the encoded coefficients via an implicit subspace partition.

To address the above-mentioned issues and \hyy{inspired by} the useful insights from traditional codecs, we make the first attempt to build a neural data-dependent transform for learned image compression.
Our new model aims to generate transform parameters dynamically based on the information of the input sample.
To this end, apart from the image content stream, our model additionally introduces \textit{neural-syntax} as the model stream to generate transform parameters at the decoder side.
The neural-syntax describes the rough contexts of images/features and therefore can make the distribution of the encoded coefficients more compact.
\cvpr{We also introduce the neural-syntax into a post-processing network which targets further enhancing the reconstruction quality when computation and time budgets are sufficient.}
With the aid of neural-syntax, our model can be online optimized towards achieving better R-D performance on the given input sample.
Similar to traditional codecs that traverse and select the best coding mode, we introduce a continuous online mode decision mechanism, optimizing the model stream codes on the input sample, to further improve the coding efficiency.
The experiments demonstrate the superiority of our method. More ablation studies and analyses show the effectiveness of each designed module as well as the rationality of our motivations and interpretations.

Our contributions are summarized as follows,
\vspace{-2mm}
\begin{itemize}
    \item We make the first attempt to build a neural data-dependent transform for learned image compression. The transform enables the decoder to be more powerful and flexible, offering superior R-D performance.
    \vspace{-1.5mm}
    \item We propose a new joint paradigm to optimize the content and model streams simultaneously, with the aid of \textit{neural-syntax} in an end-to-end image compression framework.
    \vspace{-1.5mm}
    \item The encoded coefficients of \textit{neural-syntax} are online optimized over input samples with a continuous online mode decision to further improve the coding efficiency.
\end{itemize}
}

\section{Related Work}
\subsection{Hybrid Image Compression}

\wdz{Conventional} image compression schemes follow the hybrid/transform coding paradigm.
JPEG~\cite{wallace1992jpeg} utilizes Discrete Cosine Transform (DCT) to make the 
transformed coefficients compact for bit-rate reduction and decorrelated for efficient entropy coding.
Advanced hybrid codecs, \textit{e.g.} HEVC~\cite{sullivan2012overview} and VVC~\cite{vvc}, add more types of transforms, \textit{e.g.} Discrete Sine Transform (DST)~\cite{DST_intra_pred,AT}, to handle different types of residual signals. 
Specifically, Multiple Transform Selection (MTS) is introduced in VVC to select the most desirable transform with the best rate-distortion performance.
Data-driven transforms like KLT have also been explored for image compression~\cite{zhang2020image}, where multiple KLT candidates are trained from different clusters of multi-scale patches.
Inspired by the idea of signal-dependent transform selection in previous image coding methods, our model adopts end-to-end learned neural networks to generate data-dependent transforms for more efficient image compression.
Different from existing hybrid codecs, all components in our model, including neural transforms, are trained in an end-to-end manner, and the discrete transform selection is extended into a more flexible continuous mode decision process. These two characteristics lead to more compact representations and improved R-D performance in our method.

\subsection{Deep Image Compression}
In recent years, with the surge of deep learning techniques, more attention has been paid to end-to-end image compression.
Ball{\'e} \textit{et al.} first utilized Convolutional Neural Network (CNN) to establish a compressive auto-encoder for lossy image compression~\cite{balle2015density,balle2016end} and inspired a lot of learned image compression methods \cite{minnen2017spatially,minnen2018image,cai2018deep}.
With the help of a new proposed Generalized Divisive  Normalization~(GDN), the convolutional analysis and synthesis transforms learn to reduce the redundancy of image signal effectively.
Besides the transform, numerous researches are dedicated to entropy coding on the latent representations based on the learned probability models, \textit{i.e.} hyper-prior~\cite{balle2018variational}, 2D context model~\cite{lee2018context,minnen2018joint} and 3D context model~\cite{chen2019neural} .
Besides, Gaussian Mixture Model (GMM) and attention module-based transforms~\cite{cheng2020learned} have been shown to further improve image compression performance.

Despite the evolution of learned image compression transforms and entropy models, existing methods still adopt data-independent encoding and decoding transforms.
Once the training is done, the parameters of the transforms are fixed for all input images.
Such transforms face challenges when handling diversified image signals. Wang~\textit{et al.}~\cite{ensemble} used ensemble learning to compress images with the selected models from a model pool, and achieved an improved performance.
\textcolor{black}{Model stream is also introduced in learned video coding framework \cite{van2020overfitting,van2021instance}, in which the encoder and the decoder are both overfitted on the given sequence. The updated weights are then compressed and stored in the bitstream. The overfitted model brings about significant performance gain on video sequences.}
In this work, we explore to adopt the data-dependent transforms in the learned image compression framework. 
\wdz{Beyond selecting a desired transform from a hand-crafted pool, our method dynamically generates the transform for each input in an end-to-end manner}.
\wdz{Furthermore, we generalize the discrete mode decision to a continuous version, which is more flexible to be online optimized and offers better R-D performance.}

\section{Neural Data-Dependent Transform}

\subsection{Formulation and Motivation}
\label{sec:formulation}

\begin{figure}[t]
    \centering
    \includegraphics[width=\linewidth]{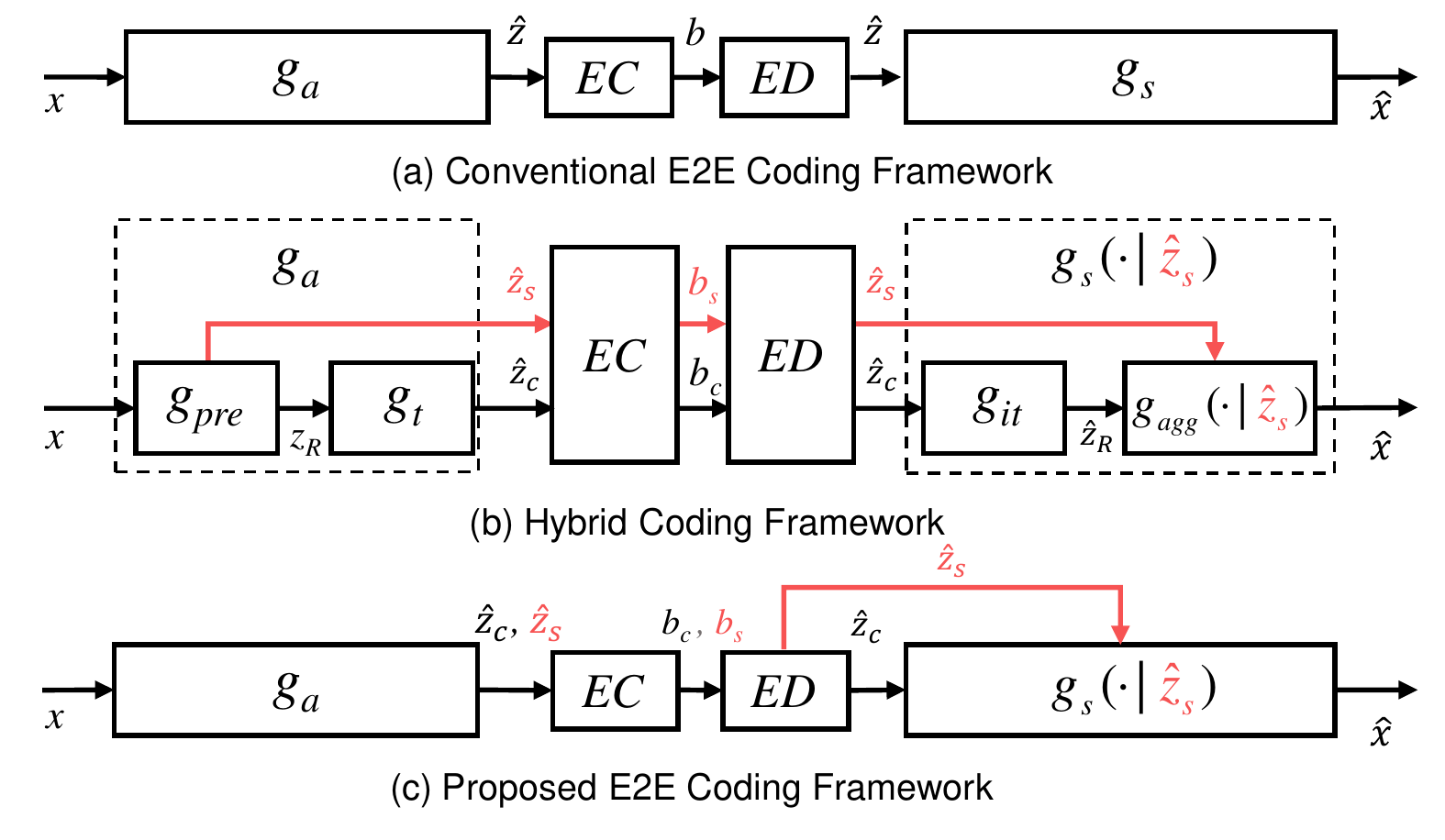}
    \caption{
    Key ideas with related formulations of image compression. 
    (a) Previous learned image compression. (b) Hybrid coding framework. 
    (c) Our dual-stream learned image compression with the data-dependent transform.
    The red lines denote the streams of syntax/neural-syntax. $g_s\left(\cdot|\hat{z}_s\right)$ denotes that the process $g_s$ is parameterized by $\hat{z}_s$.}
    \vspace{-4mm}
    \label{fig:step_by_step}
\end{figure}

In this section, we formulate several coding schemes.
With a unified formulation, the proposed model is compared with the existing end-to-end learned image compression and the hybrid coding framework to unveil our motivations clearly. 
The core idea of our formulation is illustrated in Fig.~\ref{fig:step_by_step}.
\vspace{1mm}

\noindent \textbf{Conventional Learned Compression Framework.} 
Most end-to-end learned image compression methods follow the transform coding paradigm. As shown in Fig.~\ref{fig:step_by_step} (a), the discrete latent representation $\hat{z}$ is extracted by applying a transform $g_a(\cdot)$ to an input image $x$, followed by a quantization $Q(\cdot)$ 
as follows,
\begin{equation}
    z = g_a(x),~\hat{z} = Q(z).
\end{equation}
\wdz{The quantization step in Fig.~\ref{fig:step_by_step} is omitted for brevity.}

Then $\hat{z}$ is entropy coded by an entropy coder (EC) with a predefined prior probability distribution or advanced entropy models, \textit{e.g.} hyper-prior or context models.
The bit-stream $b$ is generated by the entropy coding, and it is losslessly decoded by a symmetrical Entropy Decoder (ED). The process is denoted as follows,
\begin{equation}
    b = EC(\hat{z}),~\hat{z} = ED(b).
\end{equation}
The decoded $\hat{z}$ is then fed to a synthesis transform $g_s(\cdot)$, \textit{i.e.} \wdz{a} neural decoder, to obtain the final reconstructed output $\hat{x}$ as follows,
\begin{equation}
    \hat{x}~=~g_s(\hat{z}).
\end{equation}
\HYY{Note that this kind of compression framework is generally similar to JPEG in formulation and therefore it inevitably inherits JPEG's limitations.}
Though existing methods achieve considerably improved performance 
with the power of large-scale data and the strong modeling capacity of learned nonlinear transforms~\cite{balle2020nonlinear}, their flexibility and adaptivity might be questionable as
\HYY{it is difficult for the transforms}
to capture the specific properties of input samples.
\vspace{1mm}

\noindent\textbf{Advanced Hybrid Coding Framework.} 
Beyond the transform coding scheme, modern coding standards, \textit{e.g.} HEVC or VVC, adopt the hybrid coding scheme combining transform coding and predictive coding.
This coding scheme naturally introduces syntax elements, \textit{e.g.} intra prediction modes, partition maps, and transform modes.
Such elements are formed by the encoder by analyzing the image signal and selecting the best candidate based on R-D performance.
For simplicity, such analysis is denoted as an overall pre-processing function $g_{pre}(\cdot)$, which produces the residue component (after prediction) and the syntax component \HYY{respectively} in the whole bit-stream as follows,
\begin{equation}
    \{\hat{z}_s,z_R\}=g_{pre}(x),
\end{equation}
where $\hat{z}_s$ is the to-be-entropy-coded representation of syntax elements and $z_R$ is the to-be-transformed content information (usually residues).
Transforms $g_t(\cdot)$ \wdz{(such as DCT)} with quantization $Q(\cdot)$ are applied to $z_R$ for energy compaction, decorrelation and entropy reduction as follows,
\begin{equation}
    z_c = g_t(z_R),~\hat{z}_c=Q(z_c).
\end{equation}
To provide a uniform formulation, $g_{pre}(\cdot)$ and $g_t(\cdot)$ can be merged into an abstract integrated analysis function $g_a(\cdot)$.
The input of $g_a(\cdot)$ is the uncompressed image and the output is $\{\hat{z}_s,\hat{z}_c\}$. Therefore, we have,
\begin{equation}
    \{\hat{z}_s,\hat{z}_c\} = g_a(x).
\end{equation}
After that, both $\hat{z}_c$ and $\hat{z}_s$ are entropy coded, transmitted, and decoded in turn as follows,
\begin{equation}
    b_s=EC(\hat{z}_s),~\\
    \hat{z}_s=ED(b_s),
\end{equation}
\begin{equation}
    b_c=ED(\hat{z}_c),~\\
    \hat{z}_c=ED(b_c),
\end{equation}
where $b_c$ and $b_s$ represent the bit-stream of content information and syntax information, respectively.

\begin{figure*}[t]
    \centering    \includegraphics[width=0.75\linewidth]{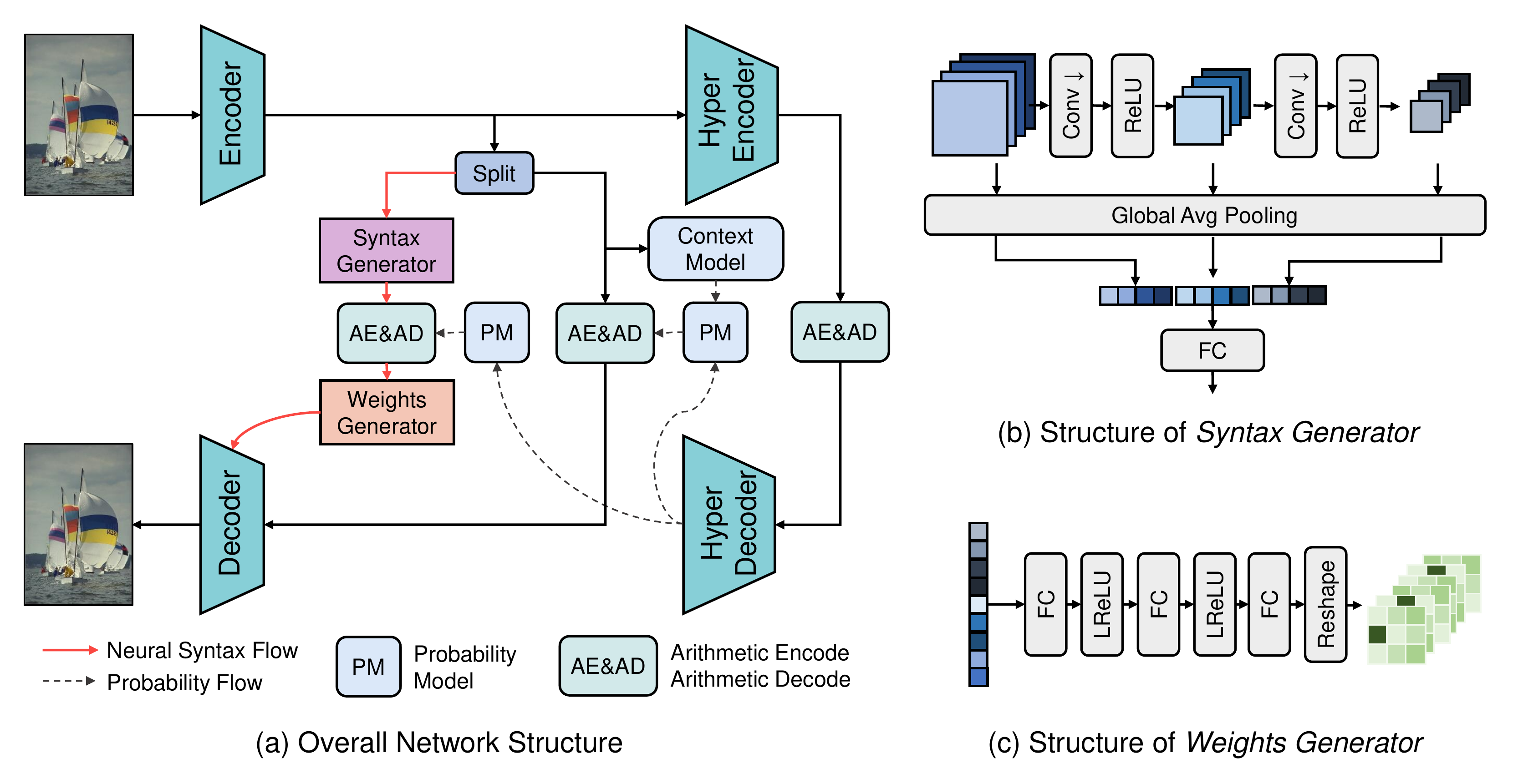}
    \vspace{-3mm}
    \caption{Pipeline and detailed network structures of the proposed dual-stream image compression framework with the neural data-dependent transform.}
    \vspace{-5mm}
    \label{fig:structure}
\end{figure*}

To reconstruct the decoded image, $\hat{z}_c$ is sent to the inverse transform $g_{it}(\cdot)$, like Inverse DCT (IDCT), to reconstruct spatial domain representation $\hat{z}_R$. The process can be denoted as,
\begin{equation}
    \hat{z}_R = g_{it}(\hat{z}_c).
\end{equation}
An aggregation component that is symmetrical to $g_{pre}(\cdot)$, denoted as $g_{agg}(\cdot)$, is applied to reconstruct the image signal.
In this step, the syntax $\hat{z}_s$ and reconstructed residue $\hat{z}_R$ are aggregated, where $\hat{z}_s$ serves as a part of parameters in $g_{agg}(\cdot)$ to control the condition of $\hat{z}_R$ to reconstruct the image. We denote the process as follows:
\begin{equation}
    \hat{x} = g_{agg}(\hat{z}_R|\hat{z}_s).
\end{equation}
Taking the decoder transform as a whole, the overall synthesis function $g_s(\cdot)$ is controlled by the syntax information to first decode and then map $\hat{z}_c$ into the reconstructed signal, as
\begin{equation}
    \hat{x} = g_s(\hat{z}_c|\hat{z}_s).
\end{equation}

The evolution of hybrid image coding naturally leads to data-dependent transforms.
However, the merits of hybrid image coding and end-to-end image learned compression are never met.
Therefore, it is much anticipated to construct a learned compression method with a neural data-dependent transform.

\vspace{1mm}

\noindent\textbf{Proposed Dual-Stream Learned Compression Framework.}
Inspired by the evolution of hybrid coding, we aim to design an image compression framework with end-to-end learned data-dependent transforms, as shown in Fig.~\ref{fig:step_by_step} (c). 
Similar to end-to-end learned frameworks, our pipeline consists of an analysis transform as the encoder, an entropy codec, and a synthesis transform as the decoder.
The transforms (modules) are all differentiable, so that they can be optimized in an end-to-end manner. 
Besides, the proposed framework differs from existing learning-based ones in the following three aspects.
\vspace{0.5mm}

\noindent $\bullet$ \textbf{\wdz{Encoding}}: the analysis transform generates not only the content representation but also neural-syntax representation as follows,
\begin{equation}
\label{eq:ourz}
    \{\hat{z}_s, \hat{z}_c\} = g_a(x),
\end{equation}
where $\hat{z}_s$ and $\hat{z}_c$ represent content information and syntax information, respectively.
The disentangled neural-syntax is similar to syntax elements in hybrid coding.
The introduced (neural-)syntax is capable of capturing
\wdz{abstract}
context information of the images/representations, which helps project the encoded representation into a subspace \HYY{where the transform coefficients are more compact}.
\vspace{0.5mm}

\noindent $\bullet$ \textbf{\wdz{Entropy Model}}: 
our entropy model differs from existing ones, where two latent representations, \textit{i.e.} $\hat{z}_s$ and $\hat{z}_c$ are separately coded. We compress them into two streams,
\begin{equation}
    b_c = EC(\hat{z}_c),~
    b_s = EC(\hat{z}_s),
\end{equation}
where $b_c$ and $b_s$ are the compressed bit-streams of $\hat{z}_c$ and $\hat{z}_s$, respectively.
The separation of the bit-stream enables finer-grained control of the coding process. For example, we can online optimize the syntax stream based on the R-D performance of the input sample.
Symmetrically, the entropy decoding is applied to the two streams,
\begin{equation}
     \hat{z}_c = ED(b_c),~
     \hat{z}_s = ED(b_s).
\end{equation}
We encode and decode $\hat{z}_s$ and $\hat{z}_c$ with the help of the hyper-prior and context model, omitted here for simplicity.
\vspace{0.5mm}

\noindent $\bullet$ \textbf{\wdz{Data-dependent Decoding}}: the proposed decoding function is data-dependent.
For different input images $x$, we obtain different neural-syntax $\hat{z}_s$ to generate a more 
\wdz{specific} 
decoding transform for the input sample. 
We denote the synthesis transform as $g_s(\cdot)$ parameterized by $\hat{z}_s$ as, 
\begin{equation}
    \hat{x} = g_s(\hat{z}_c|\hat{z}_s).
\end{equation}

Finally, we can optimize the whole pipeline in an end-to-end manner based on the rate-distortion trade-off, denoted as,
\begin{equation}
\label{eq:rdo}
    L=D(x,\hat{x})+\lambda (R(\hat{z}_c) + R(\hat{z}_s) + R(\hat{z}_h )),
\end{equation}
where $D(\cdot,\cdot)$ is the distortion metric and $R(\cdot)$ measures the bit-rate. $\hat{z}_h$ represents the hyper-prior of $\hat{z}_s$ and $\hat{z}_c$. $\lambda$ is the hyper-parameter to trade-off between rate and distortion.

\subsection{Network Design}

\subsubsection{Overall Structure}
The overall network structure is shown in Fig.~\ref{fig:structure}(a).
We adopt an end-to-end image compression framework with hyper-prior~\cite{balle2018variational} and context model~\cite{minnen2018joint,lee2018context} for entropy estimation as our baseline.
Beyond the baseline, we introduce a data-dependent transform with the help of neural-syntax/model stream, where the related stream flow is denoted by red lines in Fig.~\ref{fig:structure}(a).
Specifically, the encoder network generates the latent representations from the input image. The latent representations are split on the channel dimension to form the content stream and model stream, \textit{i.e.} neural-syntax.
The content stream, corresponding to $\hat{z}_c$ in Eq.~(\ref{eq:ourz}), is quantized and entropy coded with the estimated probability from the combination of a context model and hyper-prior.
With the combination, the estimated probability is inferred with the fused information from both already coded symbols via the context model and transmitted \hyy{hyper-prior}.
We follow existing works~\cite{balle2018variational,lee2018context,minnen2018joint,hu2020coarse} to model the likelihoods with Gaussian distributions, where the probability model generates the mean and scale of a Gaussian distribution to calculate the cumulative density function and likelihoods.
The likelihoods are directly used during encoding and decoding processes by the arithmetic coder.

For the other branch after the split, a neural-syntax generator extracts 
a compact, discrete, and one-dimension representation vector from the model stream, corresponding to $\hat{z}_s$ in Eq.~(\ref{eq:ourz}).
The neural-syntax is entropy coded with the hyper-prior based probability model.
As the neural-syntax contains no spatial information, the context model is not applied.
The decoded syntax information is fed into a weight generator network, which predicts the kernel parameters \HYY{of} the \hyy{last} convolutional layer of the decoder. This layer maps the decoded feature maps to the reconstructed image.

\subsubsection{Neural-Syntax/Model Stream}
We design a syntax generator network to leverage multi-scale redundancy to better extract the syntax information, as shown in Fig.~\ref{fig:structure}(b).
We design a multi-scale network structure to extract neural-syntax $\hat{z}_s$.
The feature maps of each scale are mapped into a one-dimension latent vector with a global average pooling operation. After that, the pooled features are concatenated together.
The mechanism of applying pooling on the feature pyramid not only makes full use of multi-scale redundancy but also leads to the scale-invariant neural-syntax.
Therefore, it enables the compression for variable-resolution images.
The latent vector is quantized and entropy coded in the entropy bottleneck, conditioned on the same hyper-prior as the latent representation of images.
After that, a multi-layer fully-connected network is utilized to map the neural-syntax representations to kernel parameters of the ultimate layer in the decoder network.
These dynamically-generated parameters improve the modeling flexibility of the network at the inference stage to adapt to diversified input images.

\begin{figure}[t]
    \centering
    \includegraphics[width=\linewidth]{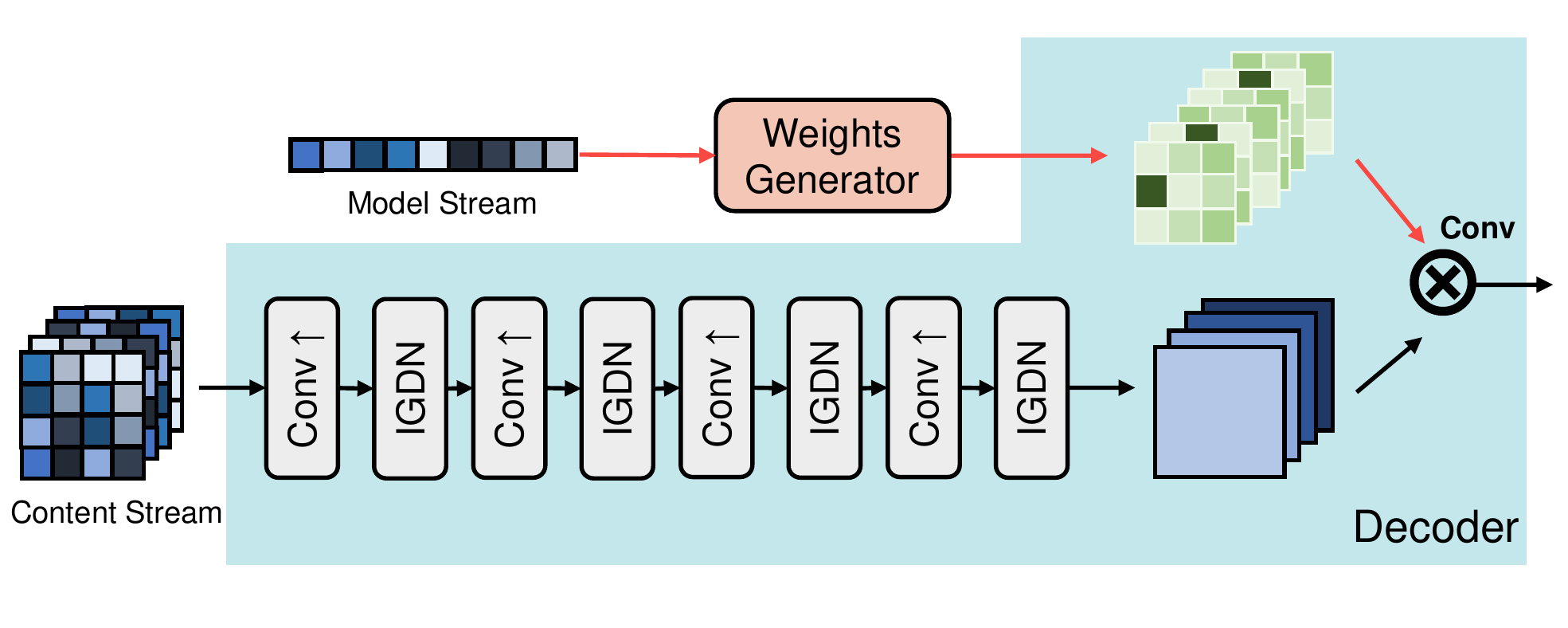}
    \vspace{-6mm}
    
    \caption{Data-dependent transform based Decoder. The transform parameters (\textit{i.e.} the convolution kernels of the last layer) are conditioned on the input image.}
    \vspace{-6mm}
    \label{fig:decoder}
\end{figure}

\begin{figure*}[t]
    \centering
\begin{subfigure}[h]{0.32\linewidth}
    \centering
    \includegraphics[width=\linewidth]{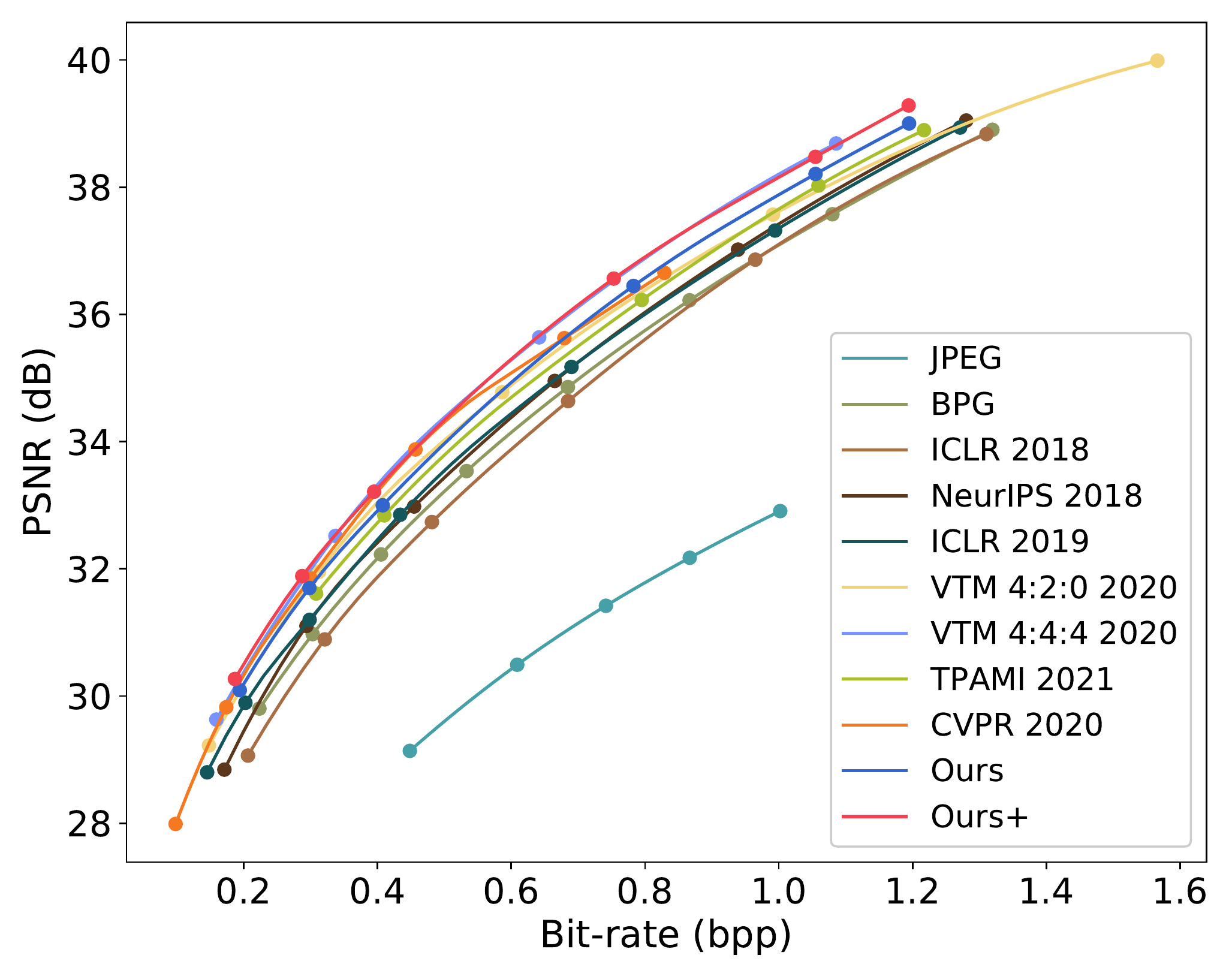}
    \caption{Kodak}
\end{subfigure}
\begin{subfigure}[h]{0.32\linewidth}
    \centering
    \includegraphics[width=\linewidth]{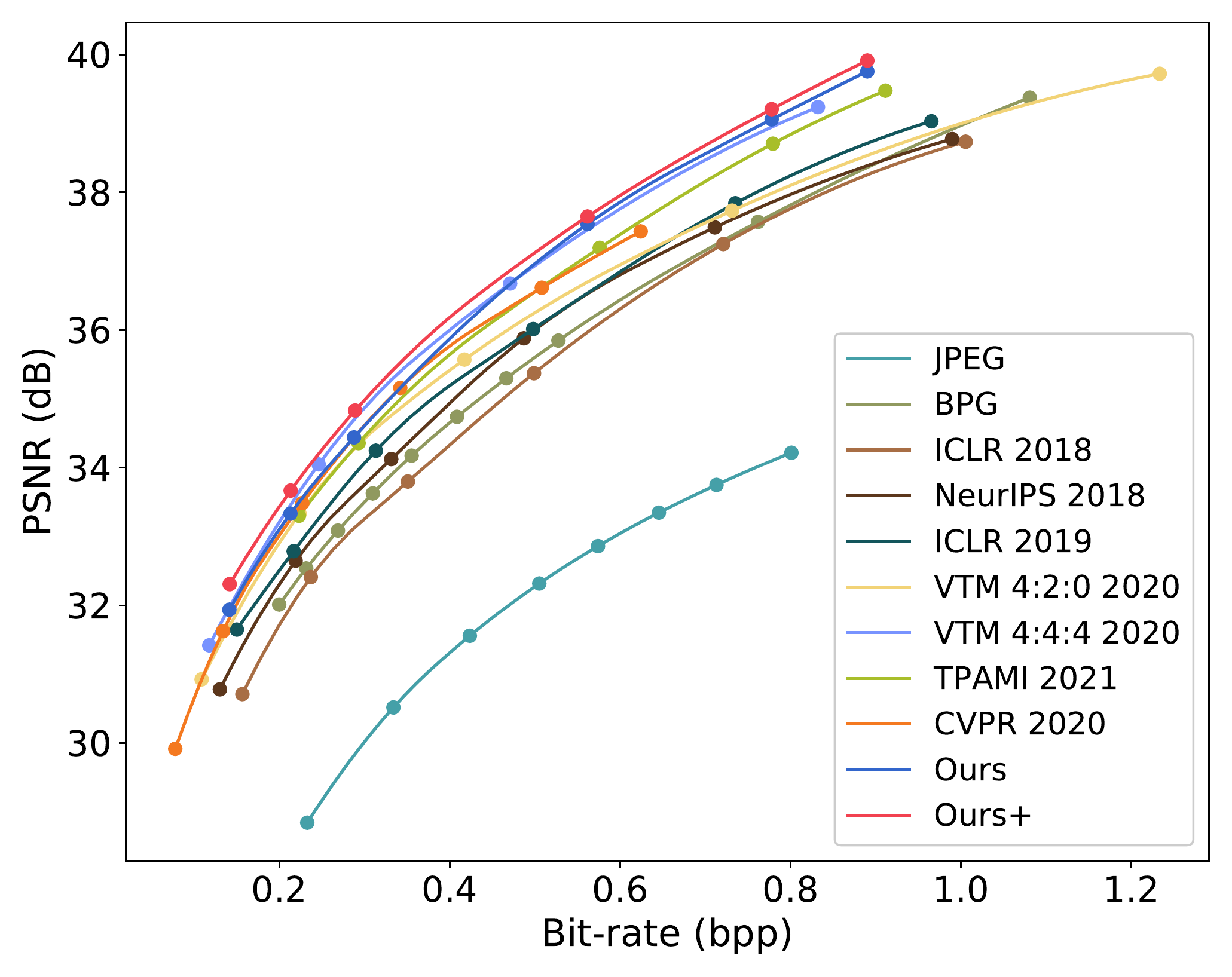}
    \caption{CLIC}
\end{subfigure}
\begin{subfigure}[h]{0.28\linewidth}
    \centering
    \includegraphics[width=\linewidth]{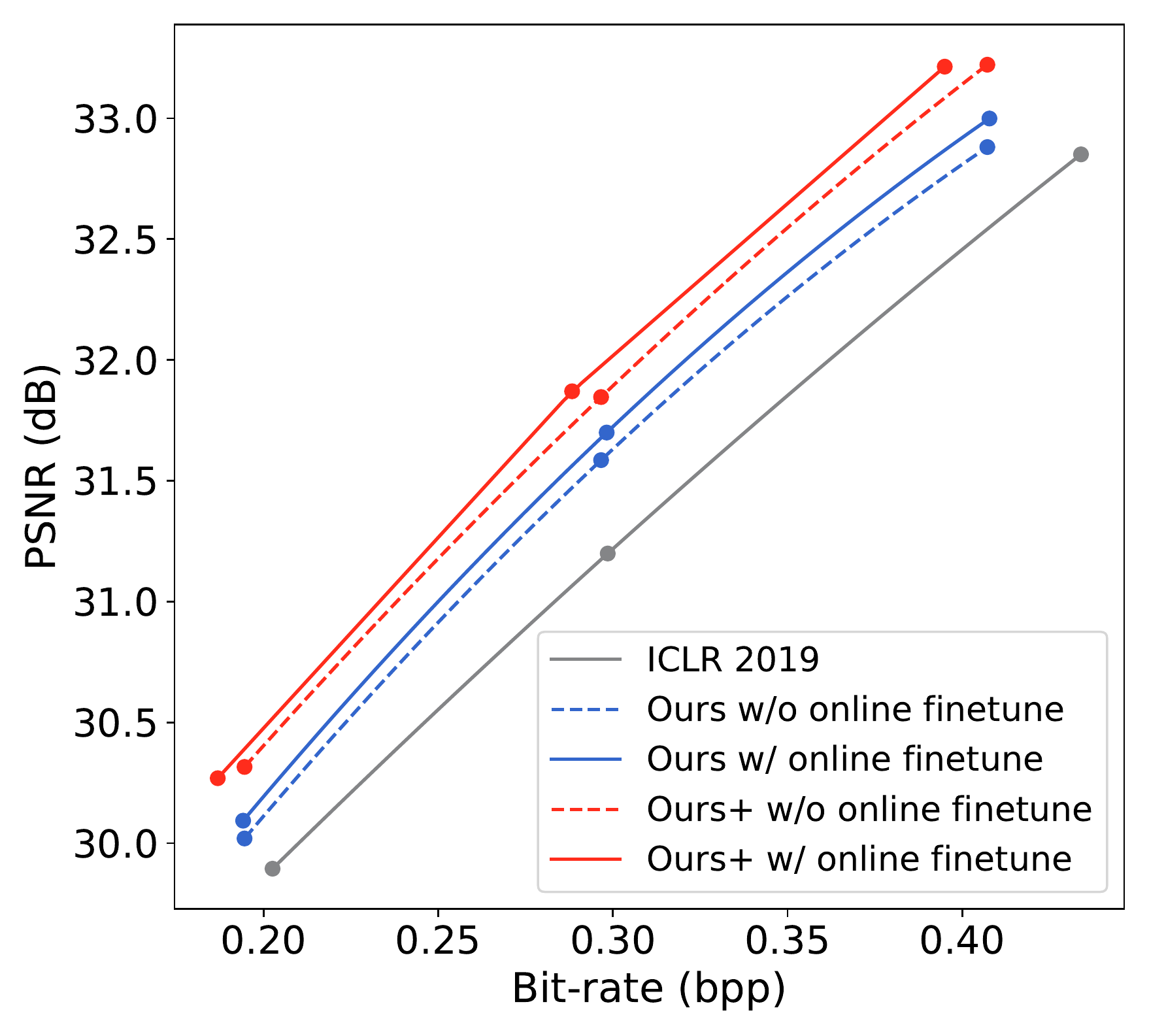}
    \caption{Ablation Studies}
\end{subfigure}
    \vspace{-2mm}
    \caption{R-D curves on Kodak and CLIC Professional Validation Set.}
    \vspace{-2mm}
    \label{fig:rdcurve}
\end{figure*}

\begin{figure*}[h]
\small
    \centering
    
    \begin{subfigure}[h]{0.15\linewidth}
        \centering
        \includegraphics[width=0.88\linewidth]{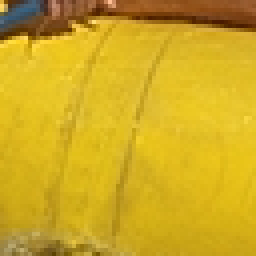}
    \end{subfigure}
    \begin{subfigure}[h]{0.15\linewidth}
        \centering
        \includegraphics[width=0.88\linewidth]{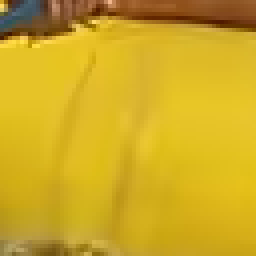}
    \end{subfigure}
    \begin{subfigure}[h]{0.15\linewidth}
        \centering
        \includegraphics[width=0.88\linewidth]{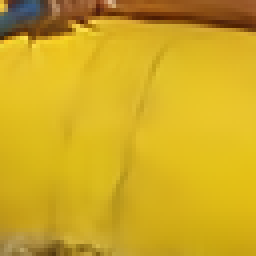}
    \end{subfigure}
    \begin{subfigure}[h]{0.15\linewidth}
        \centering
        \includegraphics[width=0.88\linewidth]{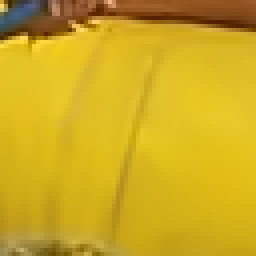}
    \end{subfigure}
    \begin{subfigure}[h]{0.15\linewidth}
        \centering
        \includegraphics[width=0.88\linewidth]{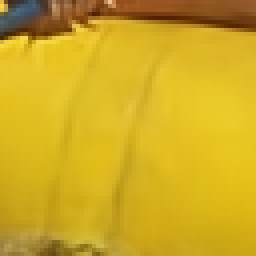}
    \end{subfigure}
    \begin{subfigure}[h]{0.15\linewidth}
        \centering
        \includegraphics[width=0.88\linewidth]{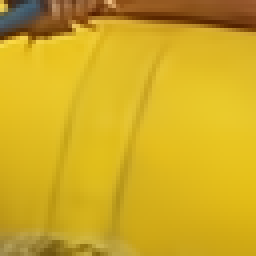}
    \end{subfigure}

    \begin{subfigure}[h]{0.01\linewidth}
    \centering
    ~
    \end{subfigure}
    
    \begin{subfigure}[h]{0.15\linewidth}
        \centering
        ~
    \end{subfigure}
    \begin{subfigure}[h]{0.15\linewidth}
        \centering
        0.555 bpp
    \end{subfigure}
    \begin{subfigure}[h]{0.15\linewidth}
        \centering
        0.582 bpp
    \end{subfigure}
    \begin{subfigure}[h]{0.15\linewidth}
        \centering
        0.586 bpp
    \end{subfigure}
    \begin{subfigure}[h]{0.15\linewidth}
        \centering
        0.549 bpp
    \end{subfigure}
    \begin{subfigure}[h]{0.15\linewidth}
        \centering
        0.531 bpp
    \end{subfigure}
    \vspace{1mm}
    \begin{subfigure}[h]{0.15\linewidth}
        \centering
        Original
    \end{subfigure}
    \begin{subfigure}[h]{0.15\linewidth}
        \centering
        BPG
    \end{subfigure}
    \begin{subfigure}[h]{0.15\linewidth}
        \centering
      ICLR 2019
    \end{subfigure}
    \begin{subfigure}[h]{0.15\linewidth}
        \centering
        \cvpr{VTM 4:4:4}
    \end{subfigure}
    \begin{subfigure}[h]{0.15\linewidth}
        \centering
        Ours
    \end{subfigure}
    \begin{subfigure}[h]{0.15\linewidth}
        \centering
        \cvpr{Ours+}
    \end{subfigure}

    \caption{Subjective results compared to BPG \cite{bpg}, ICLR 2019 \cite{lee2018context} and VTM 4:4:4 \cite{vtm8}. The patch is cropped from \textit{kodim14} in Kodak \cite{kodak}.}
    \label{fig:subjective}
    \vspace{-4mm}
\end{figure*}

The proposed neural-syntax facilitates data-dependent transforms that decode each image with different convolution kernel parameters.
The decoder is composed of five transposed convolutional layers, with the inverse-GDN in between, as shown in Fig.~\ref{fig:decoder}. 
Parameters of all but the last layer are fixed after the training process.
The last convolutional layer contains only kernel parameters without bias weights. These parameters are generated on-the-fly at the decoding time from the weight generator.
Since the parameters of the decoder are dynamically generated, they are highly dependent on the input image and can fully capture the specific properties of the input image.
Therefore, the model's adaptivity to handle diversified images is largely improved and better reconstructed images are obtained.

\subsubsection{\cvpr{Neural-Syntax based Post-processing}}
\cvpr{We also design a post-processing module for the scenarios that require higher quality reconstructions. As the parameter number of the proposed decoder is limited, it cannot fully exploit the information retained by the bitstream. To this end, we add a post-processing module with higher complexity to further enhance the original reconstructions. We adopt state-of-the-art Super-resolution methods HAN~\cite{HAN} as our backbone but remove the upsampler and replace the final convolutional layer with our dynamically generated weights. Experimental results show that by applying post-processing, the reconstruction quality can further improve by a considerable margin and outperform state-of-the-art codecs like VVC.}

\subsection{Continuous Mode Decision}
As mentioned above, the split of the bit-stream naturally brings in the potential to finetune the model/neural-syntax stream online at the inference stage.
More specifically, the decoding transform can be further optimized based on the R-D trade-off \HYY{loss function} on the input image.
The process is equivalent to the mode decision process in the conventional hybrid coding framework that selects the best one from the discrete candidates.
The online optimization of our decoder is more flexible which chooses the best one continuously from an infinite set.

It has been proposed in~\cite{eccv} to online finetune the neural network based encoder for each input image. Such an approach searches for a better representation of the image based on the fixed decoder and has been shown to achieve better performance.
However, due to the restriction that any extra updates on the decoder side should be encoded and transmitted through the channel, these methods can only adjust encoder parameters while the decoder is fixed. Thus, the degrees of freedom in finetuning are limited.

With the aid of our proposed dual-stream compression framework, the proposed approach addresses this limitation by manipulating neural-syntax in the bit-stream.
In our framework, a part of the decoder parameters are determined by the encoded representation, which can be optimized at the inference stage. The optimization is formulated as follows,
\begin{equation}
\label{eq:finetune}
\begin{split}
     \hat{\theta}_a~=\argmin_{\theta_a}&~\{D(g_s\left(g_a(x;\theta_a)|f_{syn}(x;\theta_{syn});\theta_s\right),x)\\
    &+\lambda R(g_a(x;\theta_a)) \},\nonumber
\end{split}
\end{equation}
where $x$ denotes the to-be-encoded image, and we fix the decoder parameters $\theta_s$  and syntax generator parameters $\theta_{syn}$ that are not dependent on the input.
We optimize the encoder parameters $\theta_{a}$ with R-D trade-off loss function to simultaneously search for the optimal content stream and neural-syntax stream.
It should be noted that, besides $\theta_s$, the decoder also includes generated parameters $f_{syn}(x;\theta_{syn})$, where $f_{syn}(\cdot)$ is the abstract syntax generator.
The involvement of parts of the decoder parameters provides more flexible optimization and leads to more significant performance improvement.

\section{Experimental Results}
\subsection{Implementation}
\noindent\textbf{1) Network Implementation.}
Specifically, we implement the neural-syntax model on the basis of existing end-to-end learning based image compression methods~\cite{lee2018context,minnen2018joint}, where the context model and the hyper-prior are employed. 
The detailed structure and hyper parameters of the networks are shown in the supplementary \hyy{material}. Note that the channel width of most convolutional layers is set as $N$, as well as the bottleneck width. The first $M$ channels of the bottleneck latent representation are extracted as syntax information. The remaining ($N$-$M$) channels correspond to content feature maps.
For models trained under different ranges of bit-rates, the hyper parameters are slightly changed. Models for lower ranges of bit-rates are constructed with $N = 192$ and $M = 16$.  Those for higher ranges of bit-rates have $N = 384$ and $M = 32$, to provide enough information capacity.
\cvpr{Our post-processing network is based on HAN \cite{HAN}. Specifically, we use 4 residue groups for lower bit-rate models and 6 for higher bit-rate models. We remove the upsampler and use the decoded neural-syntax to generate the weights of the final convolutional layer, which is similar to our decoder. Details of our post-processing network are also provided in the supplementary material.}

\vspace{0.5mm}

\noindent\textbf{2) Training Details.} 
We use DIV2K~\cite{Agustsson_2017_CVPR_Workshops} image set as our training dataset. The dataset is composed of 800 natural images of 2K resolutions on average. To make the model adapt to images of different resolutions, we down-sample the images to half of their resolutions as the augmentation of the training data. During training, we randomly crop $256\times 256$ patches from each image and form a batch of eight patches. 

Our training procedure consists of two stages: 1) training the compression network (including encoder, neural-syntax based decoder and entropy model) and 2) training the neural-syntax based post-processing. We fix the weights of the compression network in stage \Rmnum{2}.

\cvpr{In stage \Rmnum{1}, w}e train our models for 5,000 epochs with the Adam optimizer~\cite{kingma2014adam}. The learning rate is initialized to be 1$\times10^{-4}$ and turned down to its half after 4,000, 4,500 and 4,750 epochs. Our models are optimized by the rate-distortion trade-off loss function, defined in Eq.~(\ref{eq:rdo}). \wdz{Mean Square Error (MSE)}
\cvpr{is} used as the distortion measurement. We train \cvpr{our} models with $\lambda$ in \{$8\times10^{-4}, 1.5\times10^{-3}, 2.5\times10^{-3}, 
8\times10^{-3},
1.5\times10^{-2}, 2\times10^{-2}$\}.

\cvpr{In stage \Rmnum{2}, we train additional post-processing networks for 1500 epochs. We still adopt Adam optimizer and set the learning rate to 1$\times10^{-4}$, which is turned down to its half and quarter after 1200 and 1350 epochs, respectively. The loss function is MSE as the bit-rate will not change.}

Our method applies the continuous mode decision. For each image, based on the pre-trained network weights, we additionally employ the Adam optimizer with a learning rate 1$\times10^{-5}$ to finetune the encoder for 100 iterations. We observe a decrease in R-D loss during finetuning, corresponding to the improvement in compression performance.
\vspace{0.5mm}

\noindent\textbf{3) Evaluation Protocol.} 
We evaluate our method on Kodak image set~\cite{kodak} and the \textit{professional} subset in the CLIC validation dataset~\cite{clic}. The Kodak image set consists of 24 images, all with resolutions $768\times 512$. 
The evaluation of CLIC validation dataset reveals the performance of the proposed method on images of higher resolutions, \textit{i.e.} $1803\times 1175$ on average. 
The performance is measured by both bit-rates and distortions. We present the bit-rate in bit-per-pixel (bpp) and distortion in Peak Signal to Noise Ratio (PSNR).
The R-D curves and BD-rate~\cite{bjontegarrd2001calculation} are illustrated to compare different methods and settings.
\vspace{0.5mm}

\begin{table}[t]
  \centering
  \caption{BD-rate results ($\downarrow$) on Kodak \cite{kodak} and CLIC Professional Validation dataset \cite{clic}. We set BPG~\cite{bpg} as the anchor in the calculation. \cvpr{`Ours' and `Ours+' represent our proposed method without and with post-processing, respectively. The best results are shown in bold and the second best are underlined.}
  \vspace{-2mm}
}
    
    \footnotesize
    \begin{tabular}{lcc}
    \bottomrule
          & Kodak & CLIC \bigstrut\\
    \hline
    NeurIPS 2018 \cite{minnen2018joint} & -4.9\% & -6.2\% \bigstrut[t]\\
    ICLR 2019 \cite{lee2018context} & -5.7\% & -10.6\% \\
    VTM 4:2:0 2020 \cite{vtm8} & -9.7\% & -14.3\% \\
    VTM 4:4:4 2020 \cite{vtm8} & \textbf{-20.7\%} & \underline{-26.5\%} \\
    CVPR 2020 \cite{cheng2020learned} & -18.3\% & -22.6\% \\
    TPAMI 2021 \cite{pami_hyy} & -13.8\% & -19.5\% \bigstrut[b]\\
    \hline
    Ours  & -14.5\% & -25.3\% \bigstrut[t]\\
    Ours+ &   \underline{-20.1\%}    & \textbf{-29.7\%} \bigstrut[b]\\
    \bottomrule
    \end{tabular}%

  \label{tab:bd-rate}%
  \vspace{-3mm}
\end{table}%

\subsection{Quantitative Comparison}

We compare our method with existing end-to-end learned image compression methods optimized for MSE~\cite{balle2018variational,minnen2018joint,lee2018context,cheng2020learned,pami_hyy}\footnote{For NeurIPS 2018, we evaluate the released models based on the mean and scale hyper-prior but without the auto-regressive context model.} and conventional transform-based codecs, \textit{i.e.} JPEG~\cite{wallace1992jpeg}, BPG~\cite{bpg} and VVC~\cite{vvc}. Specifically for VVC, we use reference software VTM 8.0 \cite{vtm8} with chroma format 4:2:0 \cvpr{and 4:4:4} in the evaluation. The overall results on Kodak and CLIC Professional validation sets are shown in Fig.~\ref{fig:rdcurve} (a) and (b). We also compare the BD-rates of these methods anchored on BPG, which are shown in Table~\ref{tab:bd-rate}. \cvpr{`Ours+' in Fig.~\ref{fig:rdcurve} and Table~\ref{tab:bd-rate} represents our method with neural-syntax based post-processing. Here, we use continuous mode decision (\textit{i.e.} online finetuning strategy) for both `Ours' and `Ours+'. Results without continuous mode decision are shown in Fig.~\ref{fig:rdcurve} (c).}

\cvpr{As illustrated in Fig.~\ref{fig:rdcurve} (a) and (b), our MSE-oriented models without post-processing can already outperform recent end-to-end learned image compression methods like \cite{pami_hyy} as well as advanced conventional codecs such as VTM 4:2:0 and BPG. On CLIC, compared to the state-of-the-art learning-based method \cite{cheng2020learned}, we can even save more bit-rates at the same distortion level though we do not use attention-based transforms and Gaussian Mixture Model for entropy estimation. Once we increase the complexity of our model by applying post-processing, our method can surpass \cite{cheng2020learned} and even VTM 4:4:4. Specifically, our models save about 20.1\% BD-rate on Kodak and 29.7\% on CLIC compared to BPG. When compared to the most advanced codec VTM, our method only slightly drops 0.6\% on Kodak. While on CLIC, we can improve the BD-rate performance by about 3\%, demonstrating the effectiveness of our method.
}

\subsection{Qualitative Comparison}
We also compare our method to other codecs in visual quality. The results are shown in Fig.~\ref{fig:subjective}. 
Due to the quantization, high-frequency components in the image signal are lost in the BPG and \cvpr{VTM 4:4:4} reconstruction results. They also suffer from blocking artifacts. 
Compared to the baseline model \cite{lee2018context}, our method preserves more details. Specifically, the vertical edges in Fig.~\ref{fig:subjective} are blurred more seriously after being compressed by the baseline model, while our method reconstructs the edges better. More visual results are provided in the supplementary material.

\subsection{Ablation Study}

\noindent\textbf{1) Effectiveness of Neural-syntax.}
To verify the effectiveness of our proposed neural-syntax, we compare the proposed method with a baseline model without the neural-syntax, \textit{i.e.} the context model-based end-to-end learned compression framework~\cite{lee2018context} on Kodak.
Note that the online finetuning based continuous mode decision mechanism is not enabled in this experiment for a fair comparison. The results are illustrated in Fig.~\ref{fig:rdcurve} (c), where \textit{Ours w/o online finetune} and \textit{ICLR 2019} are compared. As shown, our model surpasses the baseline by a large margin, corresponding to 9.17\% in BD-rate. The results illustrate the effectiveness of the proposed neural-syntax mechanism.

\noindent\textbf{2) Effectiveness of Continuous Mode Decision.}
We propose to finetune the encoder together with the neural-syntax controlled decoder layer to enable continuous mode decision. 
In this experiment, we online finetune the encoder parameters for each image at the 
encoding 
time. Such finetuning makes the encoder and the decoder better adapt to the input image content.
We compare the R-D performances of the proposed models on Kodak when switching the continuous mode decision on and off.
The comparison is illustrated in Fig.~\ref{fig:rdcurve} (c), \cvpr{corresponding to \textit{Ours(+) w/o online finetune} and \textit{Ours(+) w/ online finetune}} settings. 
With continuous mode decision, the model is capable of further decreasing the BD-rate by 2.35\% on average. \wdz{It should be noted that, without online finetuning, our neural-syntax can already achieve considerable performance gain. Therefore, the additional performance improvement brought by continuous mode decision is non-trivial.}

\noindent\textbf{3) Complexity Analyses on Neural-syntax and Post-processing.} \cvpr{In previous sections, we have already demonstrated the effectiveness of the neural-syntax and post-processing. Here we further show the complexity comparison among our models (including `Ours' and `Ours+'), the baseline model \cite{lee2018context} and the state-of-the-art learning-based method \cite{cheng2020learned}\footnote{To make the comparison fair, here we use the pytorch implementation in \url{https://github.com/LiuLei95/PyTorch-Learned-Image-Compression-with-GMM-and-Attention} to align the platform.} in Table \ref{tab:complexity}. In Table \ref{tab:complexity}, we can find that our neural-syntax is rather light-weighted that only increases the parameter number by 1\%. Compared to \cite{cheng2020learned}, we use fewer parameters to achieve better performance on CLIC. And after applying the neural-syntax based post-processing, we can outperform \cite{cheng2020learned} on both Kodak and CLIC despite the fact that our parameters are still fewer.}

\begin{table}[t]
  \centering
  \caption{\cvpr{Complexity and Performance comparison among our proposed method, the baseline method \cite{lee2018context} and the state-of-the-art method \cite{cheng2020learned}. BD-rates are anchored on our baseline, \textit{i.e.}, ICLR 2019~\cite{lee2018context}}.}
  \vspace{-2mm}
    \footnotesize
    \begin{tabular}{lccc}
    \bottomrule
    \multirow{2}[2]{*}{Method} & \multirow{2}[2]{*}{\#Param} &  \multicolumn{2}{c}{BD-rate ($\downarrow$)} \bigstrut[t]\\
          &    & Kodak & CLIC \bigstrut[b]\\
    \hline
    ICLR 2019 \cite{lee2018context} & 100\% & 0\%   & 0\% \bigstrut[t]\\
    CVPR 2020 \cite{cheng2020learned} & 175\% & -12.8\% & -12.4\% \\
    Ours  & 101\% & -9.5\% & -16.5\% \\
    Ours+ & 125\% &   -15.5\%    & -21.7\% \bigstrut[b]\\
    \bottomrule
    \end{tabular}%
    \vspace{-3mm}

  \label{tab:complexity}%
\end{table}%

\cvpr{Results on MS-SSIM-oriented models can be found in our supplementary material\HYY{, where} we provide more visual comparison and ablation studies.}
\vspace{-2mm}

\section{Conclusion}
In this paper, we explore the data-dependent transforms in end-to-end learned image compression. We propose the end-to-end trained neural-syntax to provide more flexibility in the compression of diverse images. The neural-syntax mechanism also enables continuous mode decision at the inference time, allowing a further improvement in R-D performance when compressing each image. Experimental results demonstrate the effectiveness of the neural-syntax mechanism and the superior R-D performance.

{\small
\bibliographystyle{ieee_fullname}
\bibliography{egbib}

\begin{thebibliography}{10}\itemsep=-1pt

\bibitem{clic}
Workshop and challenge on learned image compression, 2020.
\newblock \url{http://www.compression.cc}.

\bibitem{Agustsson_2017_CVPR_Workshops}
Eirikur Agustsson and Radu Timofte.
\newblock {NTIRE} 2017 challenge on single image super-resolution: Dataset and
  study.
\newblock In {\em IEEE Conf. Comput. Vis. Pattern Recog. Worksh.}, 2017.

\bibitem{balle2020nonlinear}
Johannes Ball{\'e}, Philip~A Chou, David Minnen, Saurabh Singh, Nick Johnston,
  Eirikur Agustsson, Sung~Jin Hwang, and George Toderici.
\newblock Nonlinear transform coding.
\newblock {\em IEEE J. Selected Topics in Signal Processing}, 15:339--353,
  2021.

\bibitem{balle2015density}
Johannes Ball{\'e}, Valero Laparra, and Eero~P Simoncelli.
\newblock Density modeling of images using a generalized normalization
  transformation.
\newblock In {\em Int. Conf. Learn. Represent.}, 2016.

\bibitem{balle2016end}
Johannes Ball{\'e}, Valero Laparra, and Eero~P Simoncelli.
\newblock End-to-end optimized image compression.
\newblock In {\em Int. Conf. Learn. Represent.}, 2017.

\bibitem{balle2018variational}
Johannes Ball{\'e}, David Minnen, Saurabh Singh, Sung~Jin Hwang, and Nick
  Johnston.
\newblock Variational image compression with a scale hyperprior.
\newblock In {\em Int. Conf. Learn. Represent.}, 2018.

\bibitem{bjontegarrd2001calculation}
Gisle Bjontegarrd.
\newblock Calculation of average {PSNR} differences between {RD-curves}.
\newblock {\em VCEG-M33}, 2001.

\bibitem{vvc}
Benjamin Bross, Jianle Chen, and Shan Liu.
\newblock Versatile video coding.
\newblock {\em JVET-K1001}, 2018.

\bibitem{cai2018deep}
Jianrui Cai and Lei Zhang.
\newblock Deep image compression with iterative non-uniform quantization.
\newblock In {\em IEEE Int. Conf. Image Process.}, 2018.

\bibitem{vtm8}
Jianle Chen, Yan Ye, and Seung~Hwan Kim.
\newblock Versatile video coding (draft 8).
\newblock {\em JVET-Q2002-v3}, 2020.

\bibitem{chen2019neural}
Tong Chen, Haojie Liu, Zhan Ma, Qiu Shen, Xun Cao, and Yao Wang.
\newblock End-to-end learnt image compression via non-local attention
  optimization and improved context modeling.
\newblock {\em IEEE Trans. Image Process.}, 30:3179--3191, 2021.

\bibitem{cheng2020learned}
Zhengxue Cheng, Heming Sun, Masaru Takeuchi, and Jiro Katto.
\newblock Learned image compression with discretized gaussian mixture
  likelihoods and attention modules.
\newblock In {\em IEEE Conf. Comput. Vis. Pattern Recog.}, 2020.

\bibitem{bpg}
Bellard Fabrice.
\newblock {BPG} image format (http://bellard.org/bpg/). accessed: 2021-09.
\newblock 2018.

\bibitem{SVD}
Zhouye {Gu}, Weisi {Lin}, Bu sung {Lee}, and ChiewTong {Lau}.
\newblock Low-complexity video coding based on two-dimensional singular value
  decomposition.
\newblock {\em IEEE Trans. Image Process.}, 21(2):674--687, 2012.

\bibitem{hu2020coarse}
Yueyu Hu, Wenhan Yang, and Jiaying Liu.
\newblock Coarse-to-fine hyper-prior modeling for learned image compression.
\newblock In {\em AAAI}, 2020.

\bibitem{pami_hyy}
Yueyu {Hu}, Wenhan {Yang}, Zhan {Ma}, and Jiaying {Liu}.
\newblock Learning end-to-end lossy image compression: A benchmark.
\newblock {\em IEEE Trans. Pattern Anal. Mach. Intell.}, 2021.

\bibitem{kingma2014adam}
Diederik~P Kingma and Jimmy Ba.
\newblock Adam: A method for stochastic optimization.
\newblock {\em arXiv preprint:1412.6980}, 2014.

\bibitem{kodak}
E. Kodak.
\newblock Kodak lossless true color image suite (photocd pcd0992). [online].
  http://r0k.us/graphics/kodak/.
\newblock 2013.

\bibitem{Variable_block_size}
Cuiling {Lan}, Jizheng {Xu}, Wenjun {Zeng}, Guangming {Shi}, and Feng {Wu}.
\newblock Variable block-sized signal-dependent transform for video coding.
\newblock {\em IEEE Trans. Circuit Syst. Video Technol.}, 28(8):1920--1933,
  2018.

\bibitem{lee2018context}
Jooyoung Lee, Seunghyun Cho, and Seung-Kwon Beack.
\newblock Context adaptive entropy model for end-to-end optimized image
  compression.
\newblock In {\em Int. Conf. Learn. Represent.}, 2019.

\bibitem{AT}
Sung-Chang {Lim}, Dae-Yeon {Kim}, and Yung-Lyul {Lee}.
\newblock Alternative transform based on the correlation of the residual
  signal.
\newblock In {\em Congress on Image and Signal Processing}, volume~1, pages
  389--394, 2008.

\bibitem{eccv}
Guo Lu, Chunlei Cai, Xiaoyun Zhang, Li Chen, Wanli Ouyang, Dong Xu, and Zhiyong
  Gao.
\newblock Content adaptive and error propagation aware deep video compression.
\newblock In {\em Eur. Conf. Comput. Vis.}, 2020.

\bibitem{minnen2018joint}
David Minnen, Johannes Ball{\'e}, and George~D Toderici.
\newblock Joint autoregressive and hierarchical priors for learned image
  compression.
\newblock In {\em Adv. Neural Inform. Process. Syst.}, 2018.

\bibitem{minnen2017spatially}
David Minnen, George Toderici, Michele Covell, Troy Chinen, Nick Johnston, Joel
  Shor, Sung~Jin Hwang, Damien Vincent, and Saurabh Singh.
\newblock Spatially adaptive image compression using a tiled deep network.
\newblock In {\em IEEE Int. Conf. Image Process.}, 2017.

\bibitem{minnen2018image}
David Minnen, George Toderici, Saurabh Singh, Sung~Jin Hwang, and Michele
  Covell.
\newblock Image-dependent local entropy models for learned image compression.
\newblock In {\em IEEE Int. Conf. Image Process.}, 2018.

\bibitem{HAN}
Ben Niu, Weilei Wen, Wenqi Ren, Xiangde Zhang, Lianping Yang, Shuzhen Wang,
  Kaihao Zhang, Xiaochun Cao, and Haifeng Shen.
\newblock Single image super-resolution via a holistic attention network.
\newblock In {\em Eur. Conf. Comput. Vis.}, 2020.

\bibitem{Annealed}
Saurabh {Puri}, Sébastien {Lasserre}, and Patrick {Le Callet}.
\newblock Annealed learning based block transforms for hevc video coding.
\newblock In {\em ICASSP}, 2016.

\bibitem{RABBANI20023}
Majid Rabbani and Rajan Joshi.
\newblock An overview of the jpeg 2000 still image compression standard.
\newblock {\em Signal Processing: Image Communication}, 17(1):3--48, 2002.

\bibitem{DST_intra_pred}
A. {Saxena} and F.~C. {Fernandes}.
\newblock Dct/dst-based transform coding for intra prediction in image/video
  coding.
\newblock {\em IEEE Trans. on Image Processing}, 22(10):3974--3981, 2013.

\bibitem{sullivan2012overview}
Gary~J Sullivan, Jens-Rainer Ohm, Woo-Jin Han, and Thomas Wiegand.
\newblock Overview of the high efficiency video coding ({HEVC}) standard.
\newblock {\em IEEE Trans. Circuit Syst. Video Technol.}, 22(12):1649--1668,
  2012.

\bibitem{van2021instance}
Ties van Rozendaal, Johann Brehmer, Yunfan Zhang, Reza Pourreza, and Taco~S
  Cohen.
\newblock Instance-adaptive video compression: Improving neural codecs by
  training on the test set.
\newblock {\em arXiv preprint arXiv:2111.10302}, 2021.

\bibitem{van2020overfitting}
Ties van Rozendaal, Iris~AM Huijben, and Taco Cohen.
\newblock Overfitting for fun and profit: Instance-adaptive data compression.
\newblock In {\em Int. Conf. Learn. Represent.}, 2021.

\bibitem{wallace1992jpeg}
Gregory~K Wallace.
\newblock The {JPEG} still picture compression standard.
\newblock {\em IEEE Trans. on Consumer Electronics}, 38:18--34, 1992.

\bibitem{ensemble}
Yefei Wang, Dong Liu, Siwei Ma, Feng Wu, and Wen Gao.
\newblock Ensemble learning-based rate-distortion optimization for end-to-end
  image compression.
\newblock {\em IEEE Trans. Circuit Syst. Video Technol.}, 31:1193--1207, 2021.

\bibitem{mode_dependent}
Chuohao {Yeo}, Yih~Han {Tan}, Zhengguo {Li}, and Susanto {Rahardja}.
\newblock Mode-dependent transforms for coding directional intra prediction
  residuals.
\newblock {\em IEEE Trans. Circuit Syst. Video Technol.}, 22(4):545--554, 2012.

\bibitem{zhang2020image}
Xinfeng Zhang, Chao Yang, Xiaoguang Li, Shan Liu, Haitao Yang, Ioannis
  Katsavounidis, Shaw-Min Lei, and C-C~Jay Kuo.
\newblock Image coding with data-driven transforms: Methodology, performance
  and potential.
\newblock {\em IEEE Trans. Image Process.}, 29:9292--9304, 2020.

\end{thebibliography}
}

\end{document}